\documentclass[prb,twocolumn,showpacs]{revtex4}
\usepackage{graphicx}
\begin{document}

\title{Monte Carlo study  of linear chain submonolayer structures.\\
Application to Li/W(112) }
\author{ Czes{\l}aw Oleksy
\thanks{E-mail:  oleksy@ift.uni.wroc.pl}\\
Institute of Theoretical Physics, University of Wroc{\l}aw\\
Plac Maksa Borna 9, 50-204 Wroc{\l}aw, Poland
}
\date{\today}

\begin{abstract}
The lattice gas model for adsorption of alkaline elements on
W(112) surface is studied by Monte Carlo simulations. The model
includes dipole--dipole interaction as well as  long-range
indirect interaction. The numerical results show that truncation
of the indirect interaction even at 200$\AA$  may change a phase
diagram, i.e., new   phases containing  domain walls might occur.
It is demonstrated that a  defected phase can exist at high
temperatures even if it is not stable at T=0. The phase diagram
for Li/W(112) is constructed and long periodic chain structures
(9$\times 1$), (6$\times 1$), (4$\times 1$), (3$\times 1$), and
(2$\times 1$) are found to be stable at low temperatures. Role of
thermal fluctuation is discussed by comparison of Monte Carlo
results with mean field approximation results.
\end{abstract}

 \maketitle

\section{Introduction}

Chemisorption on metal surfaces has been studied for over 30 years
\cite{naumov99,persson92,zangwill,braun89,naumov77,tsong88,
pfnuer459,pfnuer62,pfnuer460}. One of interesting problems
concerns with structures and phase transitions in chemisorbed
submonolayers. In particular alkaline, alkaline--earth, and
rare-earth elements adsorbed on W(112) and Mo(112) form linear
chain structures \cite{naumov99, pfnuer459}. In case of low
coverages ($\Theta<0.5$), the commensurate phases with chains
normal to the troughs of the substrate are observed. The period
of these structures changes from 2 to 9\cite{naumov99}. At higher
coverages ($\Theta>0.5$), incommensurate surface phases occur
\cite{pfnuer62,pfnuer460}.  It has been suggested
\cite{naumov99,braun89} that formation of chain structures might
be due to indirect interaction between adatoms via conduction
electrons. Recently we proposed \cite{BLO} the lattice gas model
to account for linear chain structures on W(112) and Mo(112). The
model includes dipole--dipole interaction as well as indirect
oscillatory interaction. As the interactions are of long range,
the model is difficult to study theoretically. Preliminary
studies performed in molecular field approximation confirmed that
competition of dipole-dipole interaction and indirect interaction
leads to formation of linear chain structures. Further extensive
analysis of ground states in related effective one-dimensional
model with infinite range of interactions \cite{LO} pointed out
that competition between these two interactions is crucial in
formation of linear chain structures. In very recent
investigation of phase transitions in Li/Mo(112) and Sr/Mo(112),
H. Pfn\"{u}r et al. \cite{pfnuer459,pfnuer457,pfnuer67} have
shown that the lattice gas model could be useful in description of
long periodic phases observed at low coverages. Results of their
investigation support an idea that surface electronic state are
responsible  for the oscillatory indirect interaction between
adatoms.

It is intention of this paper to study phase diagrams by Monte
Carlo simulations. This method was successfully applied to
lattice gas models with short range interactions
\cite{selke83,binder80,binder82,binder76}. On the other hand, one
can expect more difficult problems in case of long range
interactions, e.g.,  occurrence of complicated structures
\cite{roelofs87,naumov92,bak82,sasaki94}. The paper is organized
as follows.  In section~2 the lattice gas model is described. Two
difficulties connected with application of Monte Carlo
simulations to this model are discussed in section~3.  In
section~4 phase diagrams are presented and section~5 contains
discussion and conclusions.

\section{The model}\label{s2}

\begin{figure}
\centering
\includegraphics[width=6cm]{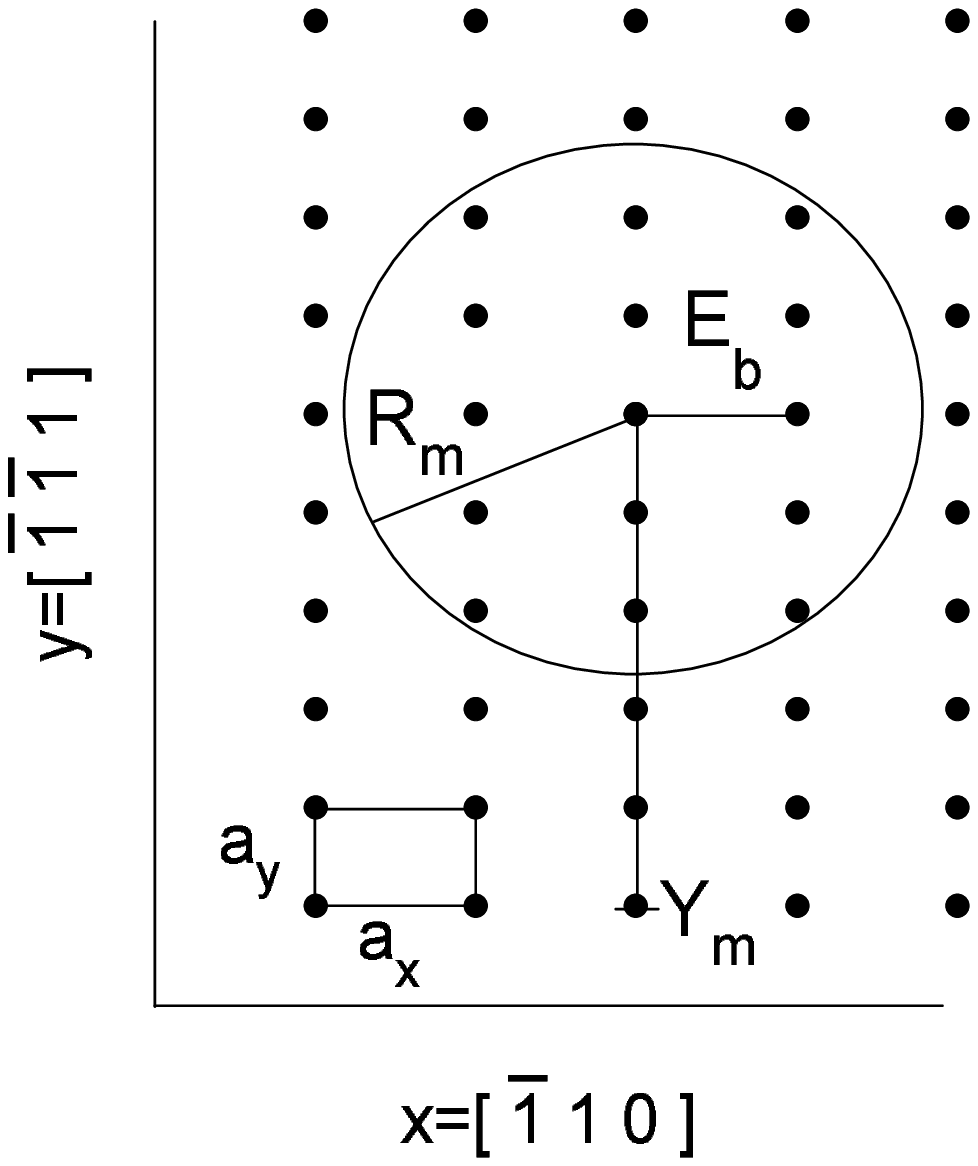}
\caption{ A lattice of adsorption sites on (112) surface of {\em
bcc} metal. The elementary cell is drawn in the left corner.
Ranges of two-particles interactions used in simulations are
depicted. For a notation see the text. }\label{fig1}
\end{figure}

In the previous paper \cite{BLO} we introduced the lattice gas
model of linear chain submonolayer structures on  the (112)
surface of W and Mo. Adsorption sites form the rectangular
lattice (see Fig.~\ref{fig1}) with $a_x=a\sqrt 2 $ and
$a_y=a\sqrt 3 /2 $, where $a$ denotes the lattice constant of the
bcc elementary cell.
 The model Hamiltonian   within the grand canonical ensemble,
is defined as
\begin{equation} \label{hamiltonian}
{\cal H} = \frac{1}{2}\sum_{i,j \atop j\neq i} V_{i j}n_{i}n_{j}-
\mu\sum_{i}n_{i} \;\;\;\;\mbox{,}
\end{equation}
where $\;n_{i}\;$ is the occupation variable  at each
adsorption site $i$ with $n_{i}=1$ if the $i$th site is
occupied by an adatom and $n_{i}=0$, if not.
The chemical potential  is  denoted by $\mu$ and
$ V_{i j} = V(\vec{r_i}-\vec{r_j})$
is the pairwise  interaction  consisting of electrostatic
and indirect interactions
\begin{equation} \label{potential}
V(\vec{r})= \frac{2d^2}{|\vec{r}|^3} + \left\{
\begin{array}{lr}
A\cos (2k_F |y|+\varphi )\frac{1}{|y|}\delta (x,0)&\;{\rm for}\;
y\neq 0\\
\\
E_b \delta (|x|,a_x)&\;{\rm for}\; y=0
\end{array} \right. \;\;\;\;\mbox{.}
\end{equation}
The first part  of Eq.~(\ref{potential}), $\; 2d^2/|\vec{r}|^3\;$,
describes a repulsive di\-po\-le-di\-po\-le interaction between
two adatoms  with identical dipole moments $d$. The second term
in Eq.~(\ref{potential}) represents an asymptotic part of the
indirect interaction between adatoms via conduction electrons of
the substrate. This particular interaction is highly anisotropic
and is closely related to the existence of nearly flattened
segments of the Fermi surface being perpendicular to the
$[\bar{1}\bar{1}1]$ axis directed along the atomic troughs of the
substrate \cite{braun89,BLO}. It is also possible that
contribution to this interaction comes from quasi-one-dimensional
surfaces states \cite{brenig95}. Very recently, G. Godzik et
al.\cite{pfnuer67} suggested that oscillatory indirect
interaction of this form is, in case of Sr/Mo(112), induced by
charge density waves involving surfaces states. An amplitude $A$
and a phase $\varphi$ can be treated as phenomenological
parameters and $k_F$ denotes a wavevector of electrons at the
Fermi surface. Finally, for $y=0$ there is an attractive indirect
interaction $\;E_b \;$ between the nearest neighbours along the
x-direction which is responsible for the chain formation.

In this work we choose the  set of model parameters the same as
proposed in the previous paper \cite{BLO}  for adsorption of
lithium on W(112) surface, i.e., $d=1.5 D$, $A=-0.137 eV$\AA,
$k_F=0.41 \AA^{-1}$, $\varphi=1.26\pi$, $E_b=-0.05 eV$, and
$a=3.16$\AA. We will use a reduced chemical potential $\mu/\mu_0$
with $\mu_0=k_B T$, $T=100K$. Although the interaction described
by Eq.~(\ref{potential}) is long-range we introduce finite range
of this interaction in Monte Carlo study. The dipole-dipole
interaction is accounted for distances between adatoms smaller
than $R_{m}$ and the indirect interaction along the y-direction
for distances smaller than $Y_{m}$ (see Fig.~\ref{fig1}).

The adatom coverage $\theta$ is defined as
\begin{equation}
\theta =\frac{1}{N}\sum_{i}\langle\: n_{i}\:\rangle,
\end{equation}
where $N$ is the number of adsorption sites and angle brackets
mean a thermodynamic average.

It is convenient to transcript the lattice gas model as Ising model by
introducing spin variables $ s_{i}= 1 - 2 n_{i}$
\begin{equation}\label{Ising}
{\cal H}= \frac{1}{8}\sum_{i,j \atop j\neq i} V_{ij} s_{i} s_{j}
-H\sum_{i}s_{i} \;\;\;\;\mbox{,}
\end{equation}
where the magnetic field $H$ is expressed by the chemical and
interaction potentials:
\begin{equation}
 H = -\frac{\mu}{2} + \frac{1}{4}\sum_{j} V_{ij}.
\end{equation}
Using Ising Hamiltonian, Eq.~(\ref{Ising}), we can perform
simulation in the canonical ensemble which is equivalent to
simulation of the lattice gas,  Eq.~(\ref{hamiltonian}), in the
grand canonical ensemble. Of course, the former method is much
easier.

\section{Monte Carlo method}\label{s3}

In Monte Carlo simulation we employ standard Metropolis algorithm
to Ising Hamiltonian, Eq.~(\ref{Ising}), on finite rectangular
lattice of $L_x \times L_y$ sites with periodic boundary
conditions (PBC). Quantities such as magnetisation $M$, specific
heat $C_{H}$, and energy $E$ are obtained as
\begin{equation}
 M = \frac{1}{N}\sum_{i} <s_{i}>,
\end{equation}

\begin{equation}
 C_{H} = \frac{1}{Nk_{B}T^{2}}\left( <{\cal H}^{2}> -
 <{\cal H}>^{2} \right),
\end{equation}

\begin{equation}
E = \frac{1}{N} <{\cal H}>,
\end{equation}
where $N=L_x  L_y$ is the number of spins. We will use the relation
between coverage  $\theta$ and the magnetisation $M$ :
$\theta = (1-M)/2$.
Since we are interested in linear chain structures we measure also
magnetisation at each row of the lattice
\begin{equation}
 M_j = \frac{1}{L_{x}}\sum_{i} <s_{{\bf r}_{i}^{(j)}}>,
\end{equation}
where
\[
{\bf r}_{i}^{(j)} = (i a_x, j a_y),
\]
Having $M_j$ one can calculate coverage at {\em j}th lattice row
$\theta_j = (1-M_j)/2$.

To study phase diagram we calculate the correlation function
$<s_j s_l>$ and the structure factor defined as \cite{selke83}
\begin{equation}
S(q) = \frac{1}{N^{2}}\sum_{j,l} <s_{j} s_{l}> \exp(i{\bf q}({\bf r}_{j}
 - {\bf r}_{l})  ),
\end{equation}
where $q$ is a wavevector from reciprocal space. This function is used
to recognize an ordered phase.

We encountered two difficulties in  examination of phase diagrams
by Monte Carlo method:
\begin{enumerate}
\item [(1)] The dependence of the results on range of the indirect
interaction.
\item[(2)] The  dependence of the resulting spin structure, at low temperatures,
 on an initial spin configuration.
\end{enumerate}
In the previous paper \cite{BLO} we studied this model within  the
molecular field approximation (MFA) finding  several ordered
phases of $(p\times 1)$ type in phase diagrams. The  $(p\times 1)$
structure at $T=0$,  has every $j p$  row (for $j=1,2,...\ldots,
L_y/p$)   occupied by adatoms and remaining rows are empty.
Therefore, coverage in this structure $\theta= 1/p$ and the
structure factor $S(q)$  reaches maximum at $q=(0,2\pi/(p a_y))$.
The phase diagram constructed for the Li/W(112) system contains
the $(2\times 1)$, $(3\times 1)$, $(4\times 1)$, $(6\times 1)$,
and $(9\times 1)$  linear chain structures. Performing Monte
Carlo simulations we observed that additional ordered structures
can occur in a phase diagram. It may happen when the range of
indirect interaction $Y_{m}$ is shorter than the linear lattice
size along the y-direction $L_y$.
\begin{figure}
\centering
\includegraphics[width=6cm]{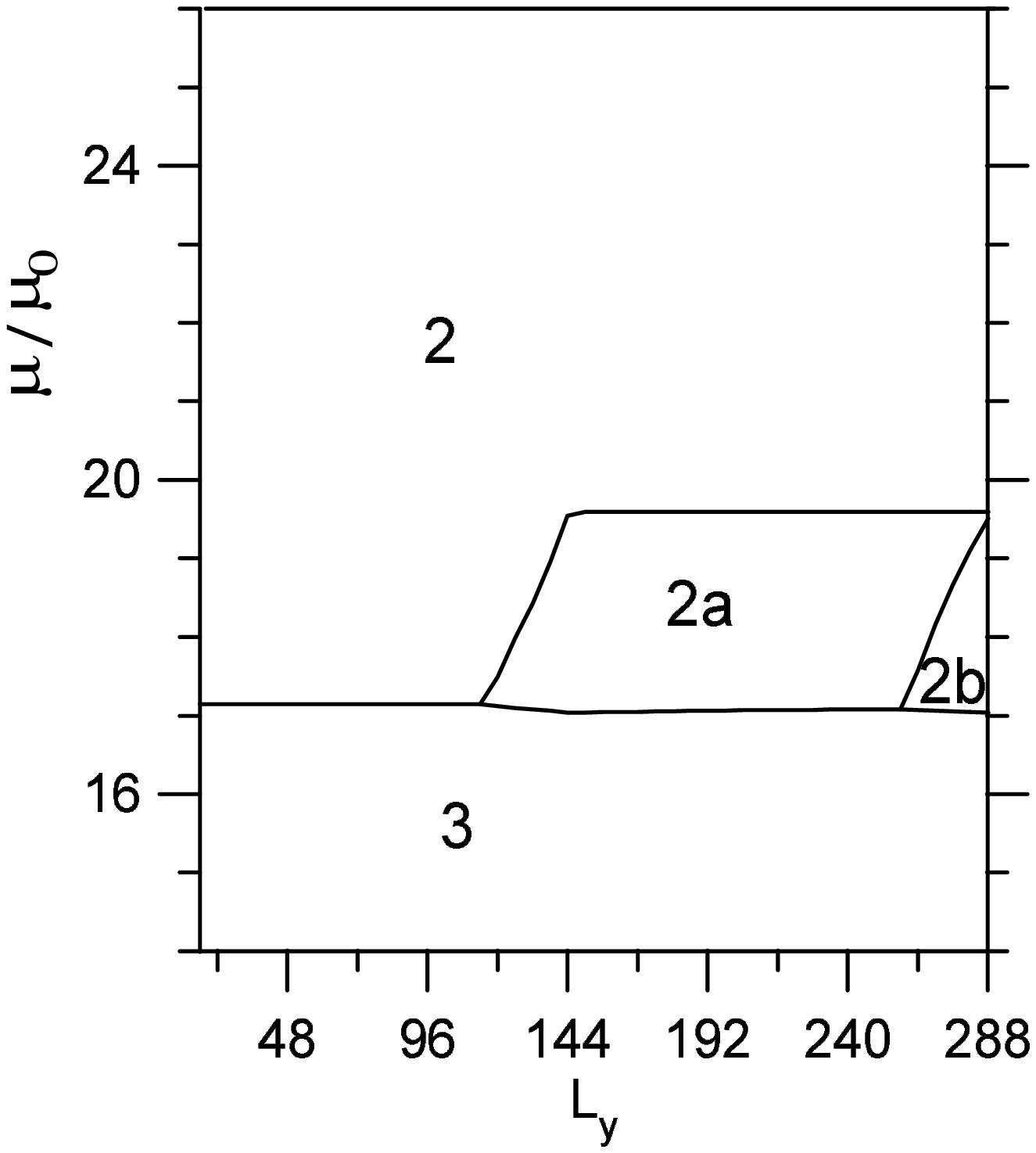}
\caption{ Ground states in the chemical potential / lattice size
$L_y$ plane for range of indirect interaction $Y_m = 72$. Regions
denoted by 2 and 3 correspond to phases ($ 2\times 1$) and ($
3\times 1$). For detailed description of states {\em 2a} and {\em
2b} see the text. }\label{fig2}
\end{figure}
We start investigation of  this problem by checking the influence
of  indirect interaction range on the ground states. As an
example we present in Fig.~\ref{fig2} a part of ground-state
diagram for constant range of interaction $Y_{m}=72$ and
different values of the lattice size $L_y$. We see that only two
phases, $(2\times 1)$ and $(3\times 1)$, are present in this part
of phase diagram when the interaction range $Y_{m}$ is close to
the lattice size ($L_y<108$). An additional phase denoted as
($2a$) appears for $L_y>108$. This is linear chain structure
consisting of two domains of the $(2\times 1)$ phase and separated
by two empty rows. Thus it has coverage $\theta= (L_y -2)/(2L_y)$.
It is worth noting that the $(2\times 1)$ phase has two domains:
one with adsorbate chains placed in even lattice rows and the
second with chains in  odd rows. The configuration of the $(2a)$
phase at $L_y=144$ may be represented by sequence of distances
between subsequent chains in the following way $\{ 2_{35}, 3_{1},
2_{34}, 3_1 \}$, where $l_j$ denotes sequence of $j$ identical
distances $l$. When range of interaction is nearly $1/4$ of $L_y$
then another phase ($2b$), containing four domain walls, occurs
in the phase diagram (see Fig.~\ref{fig2}). It looks like double
($2a$) phase, e.g., the ($2b$) phase  can be represented by
following sequences of distances $\{ 2_{34}, 3_{1}, 2_{34}, 3_1,
2_{31}, 3_{1}, 2_{27}, 3_1 \}$ for $L_y=264$. In a part of diagram
corresponding to lower coverage ( not shown in Fig.~\ref{fig2})
there are also phases containing   walls and domains of $(4\times
1)$, and $(6\times 1)$ long periodic structures  but detailed
discussion will not be presented here. It worth to mention at
this place, that the($2a$) phase can occur in the temperature
phase diagram, even if the ($2a$) phase is not a ground state. In
order to check the role of the boundary conditions, we performed
additional calculation of the ground states using free edge
boundary condition along the y-direction. The same defected
phases where obtained but the number of domain walls was reduced
by one. Above discussion shows that truncation of long-range
indirect interaction in our model causes some technical
difficulties in applying Monte Carlo method, e.g. it can be
impossible to study finite-size scaling. On the other hand, the
truncation of the interaction interpreted as screening effect
might be important from physical point of view because it
generates defected phases containing domain walls.

The second problem observed in our Monte Carlo calculation,
dependence of resulting configuration on the initial spin
configuration, is due to presence of short range attractive
interaction between the nearest neighbours along the x-direction.
When linear chain structure is formed at low temperature, it is
difficult to change the period of such structure  by changing
temperature or chemical potential using single spin flip
Metropolis algorithm. We study a phase diagram performing MC runs
at constant magnetic field (or chemical potential) by cooling
down or heating up the sample. The simulation start from high
temperature in the cooling down process and after thermalization
and measurement temperature is lowering the  by $\Delta T$ then
such procedure is repeated till $T=0$. At each temperature,  all
quantities are measured after $2 \times 10^4$ MC steps per spin
during subsequent 5000 MC steps. The final spin configuration  at
$T$ is used as initial one at $T-\Delta T$. For some values of H
the cooling down process finishes at $T=0$ with configuration
having energy greater than energy of ground-state. Hence this way
can produce metastable phases. To overcome this problem we use
heating up process which starts at temperature $\Delta T$ with
ground-state configuration as the initial spin configuration. In
subsequent  temperature steps, $T_i = T_{i-1}+ \Delta T$, the
specific heat $C_H$ and the mean energy $<{\cal H}>$ are
measured. This allow us to calculate the entropy
\[
S(T) = \int_{0}^{T}{ \frac{C_{H}(T')}{T'} dT'},
\]
by numerical integration. Then the free energy can be obtained as
\[
F(T) = <{\cal H}> -TS(T).
\]
Repeating simulations at given $H$ with different ground-state
configurations as initial configurations, we can find  a phase
with the lowest free energy at  temperature T.

\section{Phase diagrams}\label{s4}

To study phase diagrams by Monte Carlo simulations we used method
described in the previous section, i.e., simulations at constant
chemical potential (or magnetic field) with variable temperature.
The phase transition between ordered phases is determined by
examination of free energies of appropriate phases. Temperature
of transition from disordered phase to an ordered one is
determined from behaviour of the structure factor $S(q)$,
specific heat $C_H$ or sublattice magnetisation $M_j$. Lattice
sizes $L_x$, $L_y$ used in  our simulations are rather small due
to long-range interaction and large number of MC runs needed to
construct phase diagrams. Most  simulations were performed for
$20\times 72$ lattice sites, but some calculations were performed
on $40\times 72$ and $20\times 144$. Periodic boundary conditions
were assumed, therefore $L_y$ was chosen as multiplicity of 36  in
order to allow formation of ideal chain structures with period 2,
3, 4, 6, and 9, which exist in the  ground-state phase diagram.

\begin{figure}
\centering
\includegraphics[width=6cm]{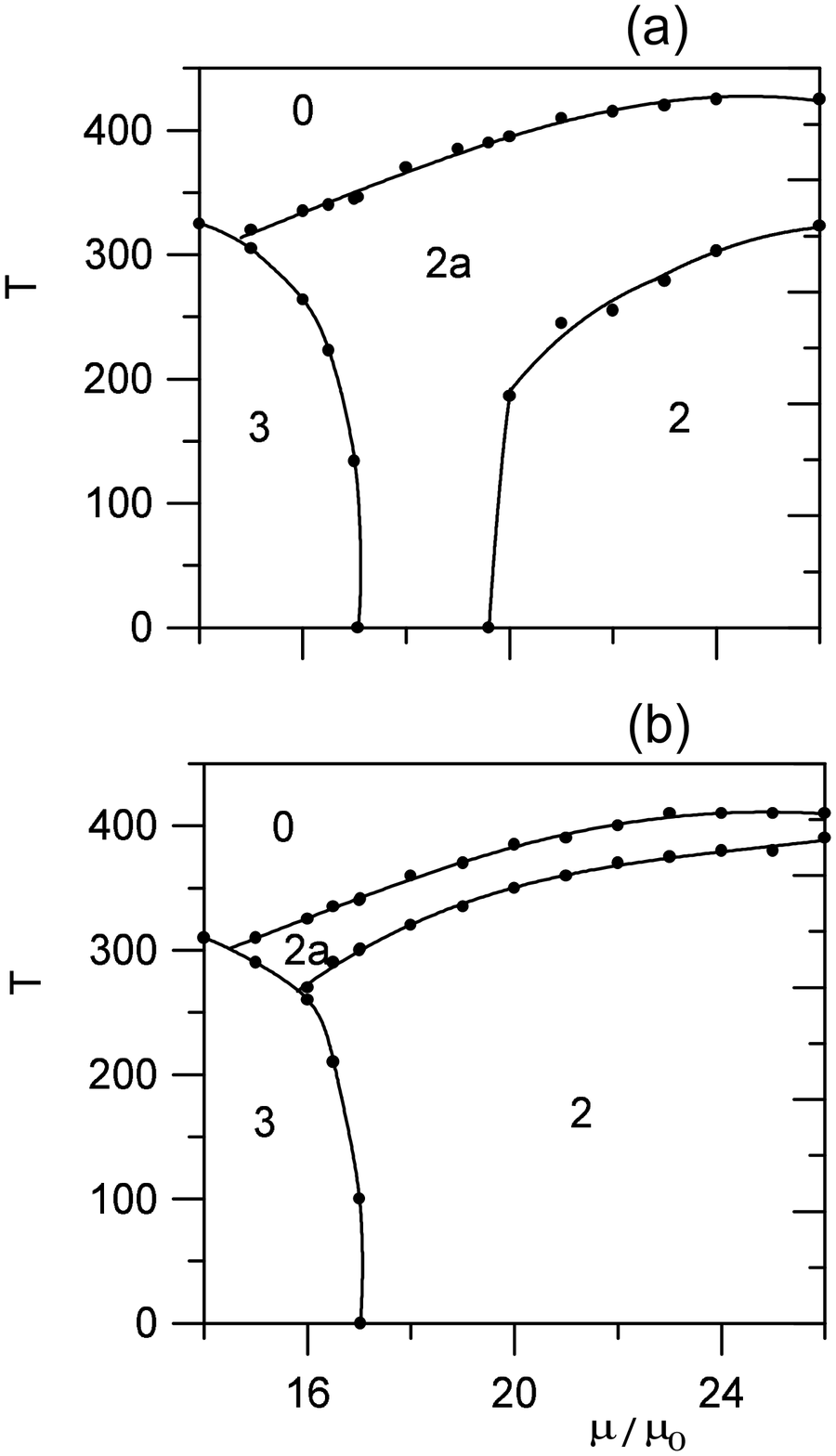}
\caption{ Phase diagrams in the temperature / the chemical
potential plane for (a) $Y_m=72$,  $L_y=144$, (b) $Y_m=36$,
$L_y=72$. Monte Carlo results are represented by circles.
}\label{fig3}
\end{figure}

\begin{figure}[h]
\centering
\includegraphics[width=6cm]{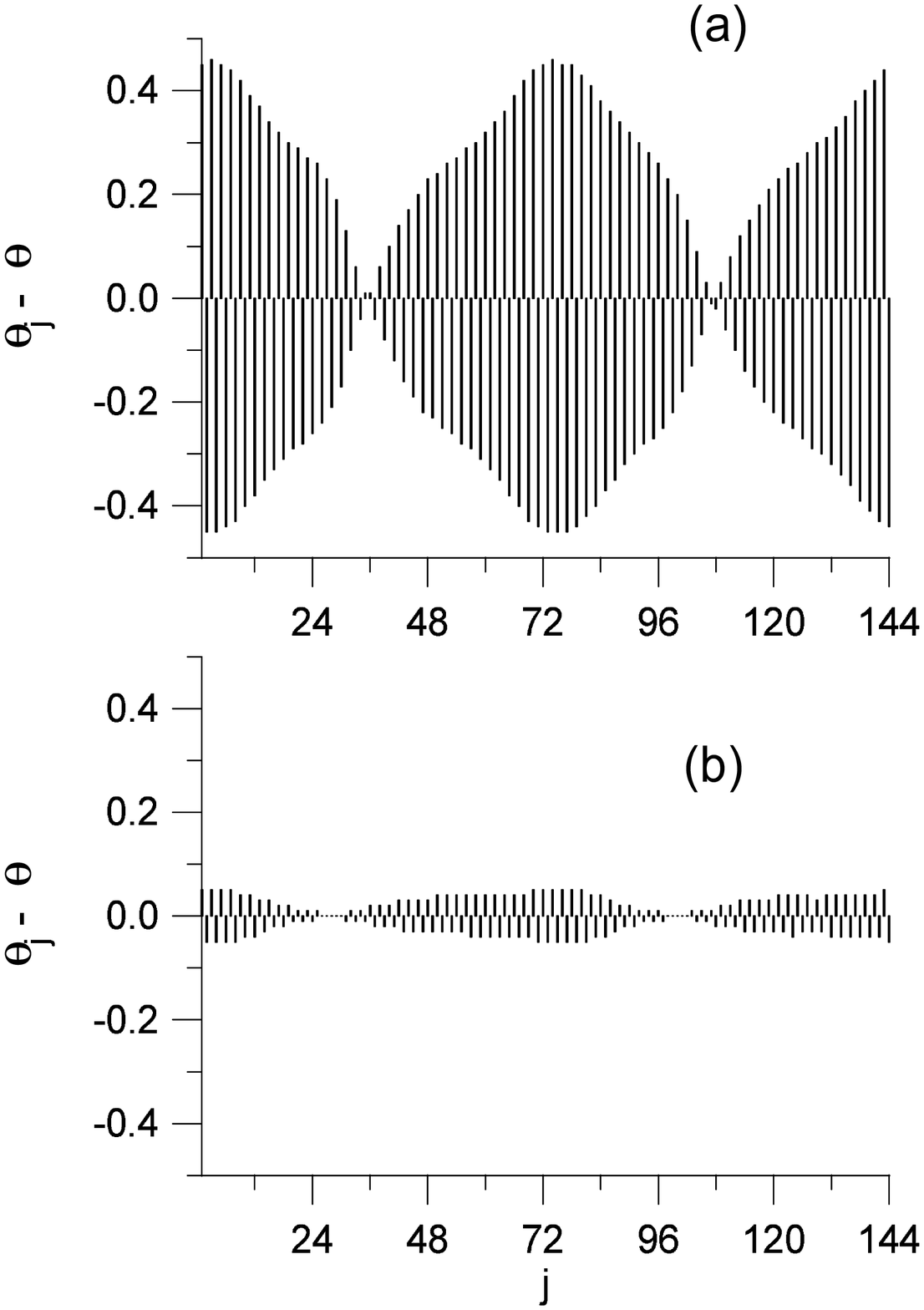}
\caption{ The deviation of  row coverage $\theta_j$ from the
lattice coverage  $\theta$ (denoted by bars) calculated at $T =
300 K$, $\mu= 22 \mu_0$, $Y_m=72$,  $L_y=144$, and  for averaging
time   (a) $5\times 10^4$  (b) $ 10^6$ MC steps per spin.
}\label{fig4}
\end{figure}

It has been shown in the previous section that defected ordered
structure can occur when $Y_{m} < L_y$. We now discuss the
thermal stability of such structure for two sets of parameters
$L_y$ and $Y_{m}$. We see in Fig.~\ref{fig3}a that the ($2a$)
phase dominates the part of phase diagram corresponding to
intermediate coverage and it pushes down the phase $(2\times 1)$.
Hence, it is impossible to pass directly  from the ordered
$(2\times 1)$ phase to the disordered phase. More interesting
case is shown in Fig.~\ref{fig3}b where the ($2a$) phase occurs
at high temperatures, although it is not stable at T=0. In this
case, the ($2a$) phase is also placed  between  the $(2\times 1)$
phase and the disordered phase. We think, that this is important
result because it shows that analysis of the ground states is not
sufficient to find all ordered phases. The phase ($2a$) has two
interesting features
\begin{enumerate}
\item [(1)] The wavevector  $q = (0,2\pi (L_y
-2)/(2L_ya_y)$, at which the structure factor  has maximum, does
not depend on temperature.
\item[(2)] The sublattice order parameters vanish  at some temperatures
whereas  the behaviour of $S(q)$ indicates the existence of
ordered structure.
\end{enumerate}
The first feature indicates that the  phase ($2a$) is not an
incommensurate phase. In order to understand the second feature
we performed simulations with different Monte Carlo measurement
times. If the time of measurement is not too long (e.g.,
$5\times10^4$ MC time steps at $T=300 K$) we observe  (see
Fig.~\ref{fig4}a) that order in this structure changes spatially
along y-direction.
The disorder appears near domain walls but as one moves form the
wall the chain structure emerges with increasing ordering. On the
other hand we found that  domain walls and the whole structure are
moving along the y direction during simulation. This is a reason
why the sublattice order parameters vanish  if averaging time is
long enough (see Fig.~\ref{fig4} where results of simulations with
different averaging time are presented) but  the structure factor
remains finite.

\subsection{Application to Li/W(112)}\label{ss4}

Linear chain structures $(2\times 1)$,  $(3\times 1)$, and
$(4\times 1)$ were observed in the experiment of adsorption of
lithium atoms on W(112) surface \cite{naumov73}. Using MFA
method, we have shown\cite{BLO} that lattice gas  model,
Eq.~(\ref{hamiltonian}), describes such sequence of structures.
Now we present results of Monte Carlo simulations. To avoid
occurrence of phases with domain walls of ($2a$) type, we chose
the range of indirect interaction equal to the  linear size of
the lattice along y direction, i.e., $Y_{m}=L_y$. Most simulations
were performed on $20\times 72$ lattice and sometimes the lattice
of $40\times 72$ sites was used. The range of dipole-dipole
interaction was $R_m=12 a_y$. The ranges of these interactions
used in previous MFA studies \cite{BLO} were both equal to
$36a_y$. Therefore we also repeated the MFA calculations for
present values of interaction ranges in order to compare the MFA
results with Monte Carlo simulation results.
\begin{figure}
\centering
\includegraphics[width=6cm]{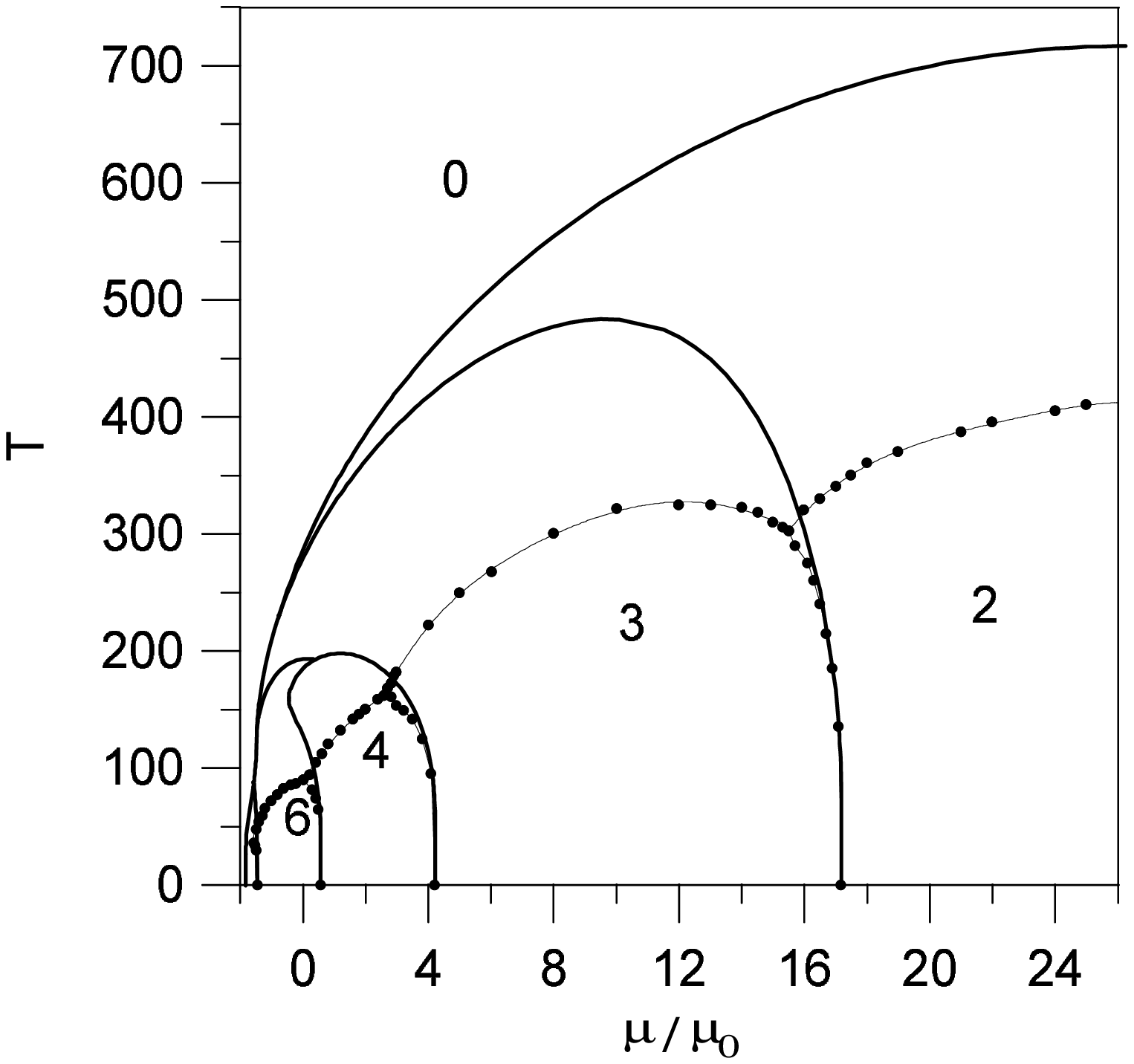}
\caption{ Phase diagram in the temperature / the chemical
potential plane for $Y_m=L_y=72$. Mean--field results are denoted
by thick lines and Monte Carlo results are represented by circles
(a thin line serve as guide for an eye). }\label{fig5}
\end{figure}
As previously, MFA calculation were performed for  all structures
of ($p\times 1$) type with $p$ up to 12. In Fig.~\ref{fig5} we
plot results of MFA and MC methods in the (T,$\mu$) phase
diagram. A low temperature part of the diagram does not depend on
the method. However in the upper part of the diagram situation is
quite different. Thermal fluctuations, not present in MFA, destroy
ordered linear chains structures at temperatures much smaller
than MFA critical temperature (the MFA critical temperature for
the ($2\times 1$) phase  can be even two times greater!).
Moreover, MFA results do not allow to reach long periodic phases
$(4\times 1)$ and $(6\times 1)$ directly from the disordered
phase. On the other hand results of MC simulations show that  all
ordered phases can be reached from disordered phase  in wide
range of chemical potential. It is interesting to look at results
of MC study presented in the (temperature - coverage) phase
diagram (see Fig.~\ref{fig6}).
\begin{figure}
\centering
\includegraphics[width=6cm]{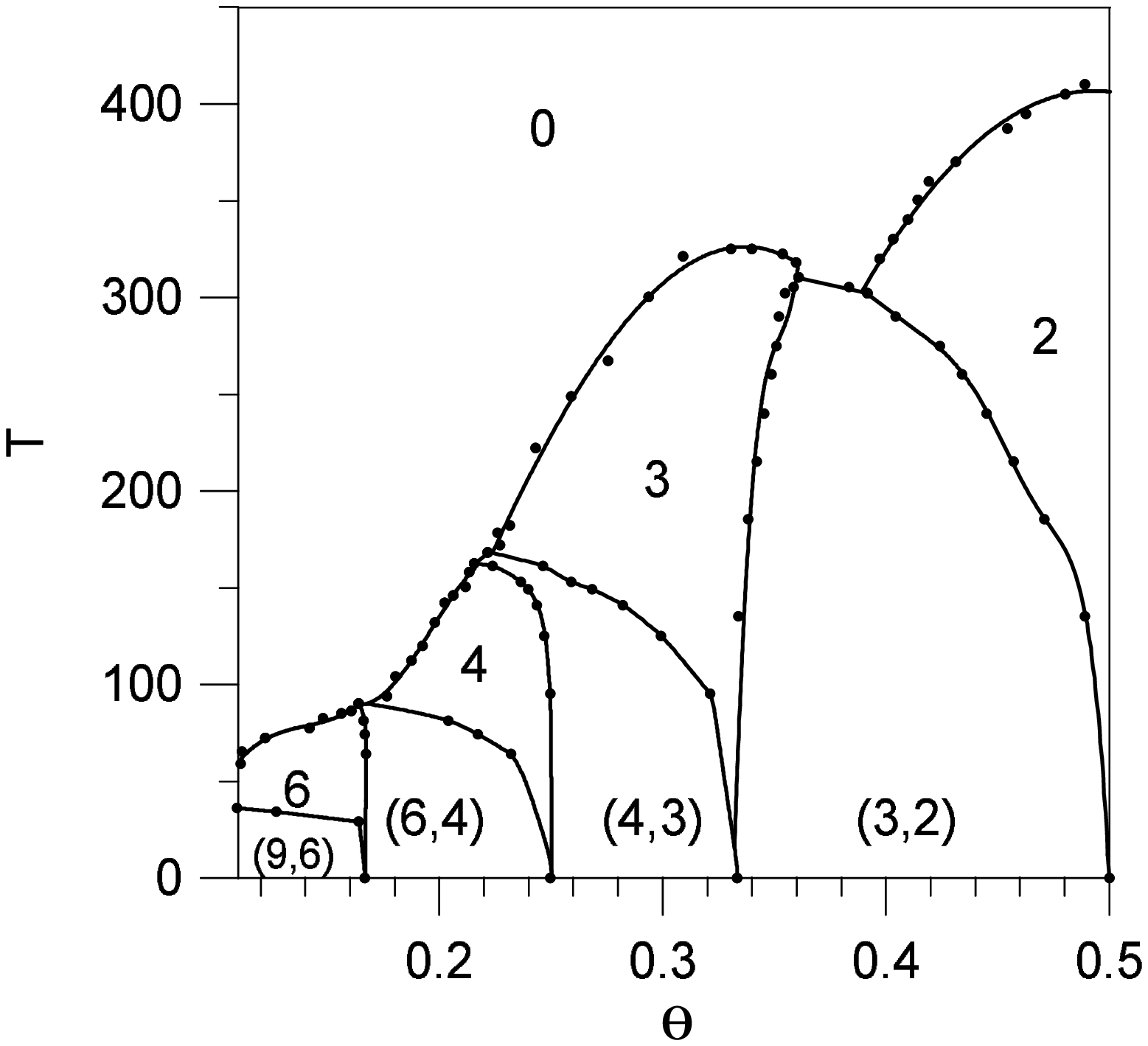}
\caption{ Monte Carlo phase diagram in the temperature / coverage
plane for $Y_m=L_y=72$. Results of simulations are represented by
circles. }\label{fig6}
\end{figure}
Ordered phases of ($p\times 1)$ are separated by mixed phases
$(p,p^{\prime})$ which denotes mixture of ($p\times 1)$ phase and
($p^{\prime} \times 1)$ phase. It is worth noting that  ordered
chain structures with period 6 and 9, present in the phase
diagram, were not observed in experiments\cite{naumov73} which
were performed at temperatures above 77K. We expect that these
phases could be observed experimentally at temperatures below 77K.

\begin{figure}
\centering
\includegraphics[width=6cm]{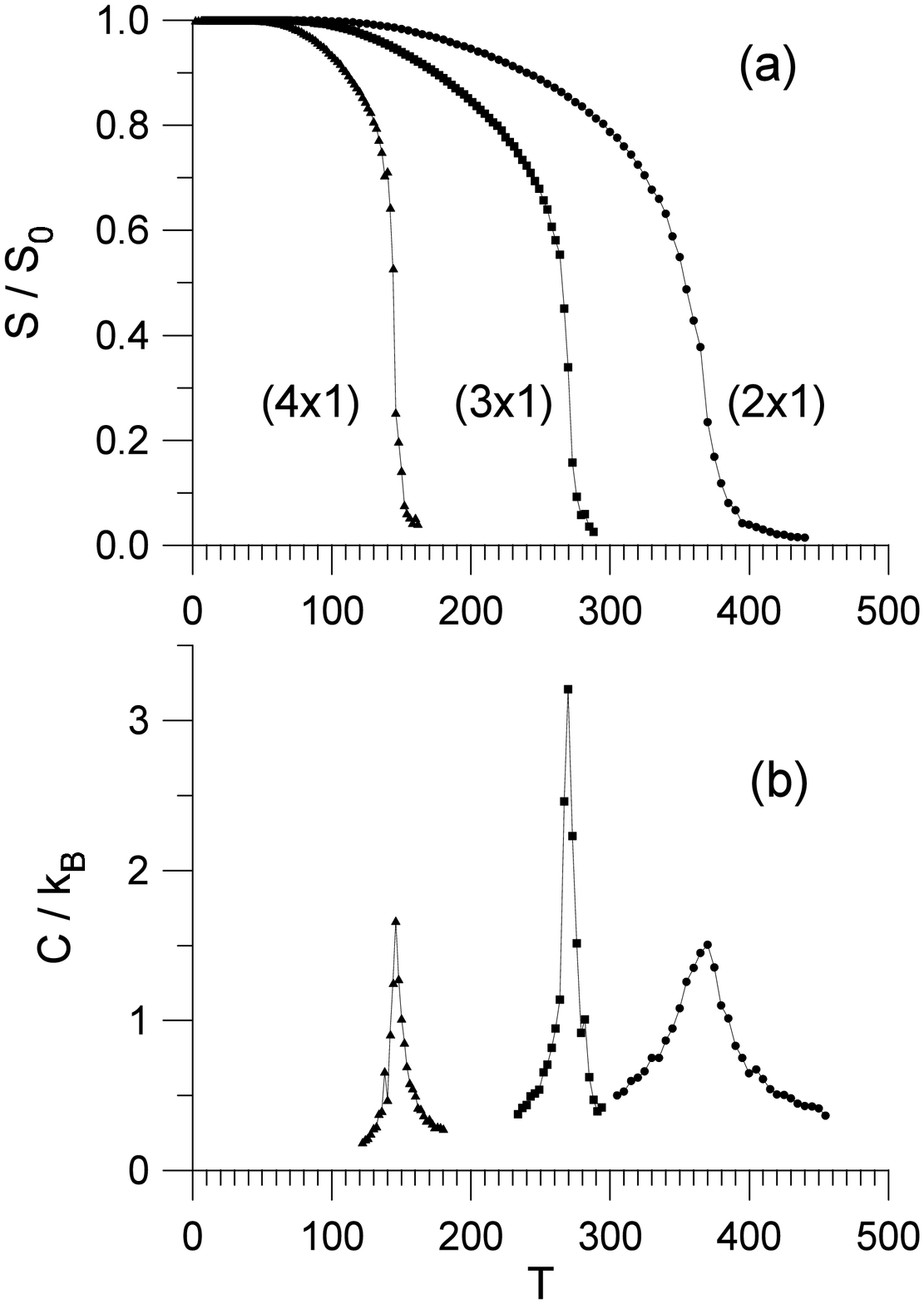}
\caption{ Temperature dependence of (a) the structure factor (b)
heat capacity in $(4\times 1)$, $(3\times 1)$, and $(2\times 1)$
phases for $\mu/\mu_0=$ 1.8, 6, and 19, respectively.
}\label{fig7}
\end{figure}

Now, we are going to discuss the order of the phase transitions.
The phase transitions between the ordered phases, e.g., $(2\times
1)$ $\rightarrow$ $(3\times 1)$, are of the first order because
sublattice order parameters and structure factors change
discontinuously at the transition points. The analysis of the
phase transition from an ordered phase to the disordered phase is
much more complicated. Checking temperature dependence of the
structure factor and heat capacity (some results are shown in
Fig.~\ref{fig7})
we found that the transition to $(9\times 1)$ and $(6\times 1)$
phases is the first-order phase transition because of
discontinuity of the structure factor. The  transition to  the
$(3\times 1)$ and $(4\times 1)$  phases seems to be continuous
transition. It is difficult to state the order of the phase
transition form the disordered phase to the $(2\times 1)$ phase.
Although, the structure factor seems to be continuous function of
temperature, the heat capacity is smeared near the transition
temperature indicating the first-order phase transition. On the
oder hand, in the MFA calculation this phase transition is of the
second order.  One of possible explanation of this discrepancy is
based on observation of additional, very weak maxima of the
structure factor with the  wave vector close to $q=(0,\pi/(a_y))$
and temperature in the narrow range of the transition temperature.
This can suggest that defected phases like the 2(a) phase are
still present close to the transition temperature. In order to
answer whether this problem is connected with the finite range of
indirect interaction one should performed finite-size calculation
with infinite range of the interaction. It can be done by
employing a numerical method to accelerate the convergence of a
Fourier series \cite{oly96}.

\section{Discussion and conclusions}
\label{s5}

Our Monte Carlo analysis revealed that truncation of the indirect
interaction  changes a phase diagram even if  the interaction
range  is very large. New phases  containing domain walls occur
at high temperatures between the disordered phase and the
$(p\times 1)$ phase. This problem complicates study of critical
behaviour via finite-size scaling. One possible solution of such
problem is accounting for infinite range of indirect
interaction.  It can be carried out by employing a numerical
method to accelerate the convergence of a Fourier
series\cite{oly96} in the same way as in earlier paper \cite{LO}.
However, it would be time consuming way to study the critical
behaviour by Monte Carlo simulations. On the other hand it would
be interesting to extend an analysis of defected phases (e.g.,
($2a$)  phase) generated by truncation of indirect interaction.
The structure factor of the  ($2a$) phase has maximum at the
wavevector slightly shifted from the position which corresponds
to maximum  in the ordered $(2\times 1)$ phase. Hence it might be
important in analysis of  LEED patterns.

Present MC calculations demonstrate that results of the mean
field approximation are very bad especially at high temperatures.
We have shown that MFA  transition temperature from  the
disordered phase to the $(2\times 1)$ phase is almost two times
greater. Moreover, the high temperature part of the phase diagram
has different topology than that  obtained by MC simulations.
However, it is interesting to note that both method give very
similar results at low temperatures. Present MC calculations
confirm earlier obtained results \cite{BLO} that at low
coverages  of lithium on the W(112) the long periodic structures
($9\times 1)$ and ($6\times 1)$ are stable at temperatures below
77K. It would be very interesting to determine experimentally the
whole (T,$\Theta$) phase diagram to check theoretical prediction.
We are aware that present lattice gas model contains some
phenomenological parameters, e.g., short range interaction $E_b$
which has influence on the transition temperature. We think that
the better way to estimate the model parameters is to perform the
first-principles calculations of the interaction potentials in the
Li/W(112) system.

Recently Yakovkin studied \cite{yakov93} chain structures
adsorbed on Mo(112) and Re(1010)  using lattice gas model in
which the indirect interaction  takes nonzero value only for one
from all lattice distances. Performing MC simulations in canonical
ensemble he was able to reconstruct the sequence of phases
($4\times 1$) and ($2\times 1$) as observed experimentally in
Li/Mo(112). It is interesting whether such simple model could
describe correctly the phases observed in adsorption of lithium
on W(112).

We observed that Monte Carlo study of phase diagram in canonical
ensemble is very difficult due to dependence of final MC results
on initial lattice gas configuration.  Although similar
difficulty  appears  in grand canonical ensemble it is rather
easy to find ground-state configurations and calculate free
energies for different phases. However, it would be interesting
to look for different algorithm than single spin flip in order to
overcome the dependence MC results on initial configurations.

\end{document}